\documentclass[prl,preprint,a4paper]{revtex4}
\usepackage{psfig,epsf}
\usepackage{graphicx}
\begin{document}

\title{Time and energy-resolved two photon-photoemission of the 
Cu(100) and Cu(111) metal surfaces }

\author{Daniele Varsano, M.~A.~L. Marques and  Angel Rubio}
\affiliation{Departamento de F\'{\i}sica de Materiales, Facultad de Qu\'{\i}micas,
Universidad del Pa\'{\i}s Vasco and Donostia International Physics Center (DIPC),
20080 San Sebastian (Spain)}

\begin{abstract}
We present calculations on energy- and time-resolved two-photon 
photoemission spectra of images states in Cu(100) and Cu(111) surfaces.
The surface is modeled by a 1D effective potential and the states are
propagated within a real-space, real-time method. To obtain the energy 
resolved spectra we employ a geometrical approach based on a subdivision
of space into two regions. We treat electronic inelastic
effects by taking into account the scattering rates calculated within a 
GW scheme. To get further insight into the decaying mechanism we have 
also studied the effect of the variation of the classical Hartree potential 
during the excitation. This effect turns out to be small.
\end{abstract}

\maketitle

\section{Introduction}
\begin{figure}
\begin{center}
\includegraphics[width=0.75\textwidth,clip]{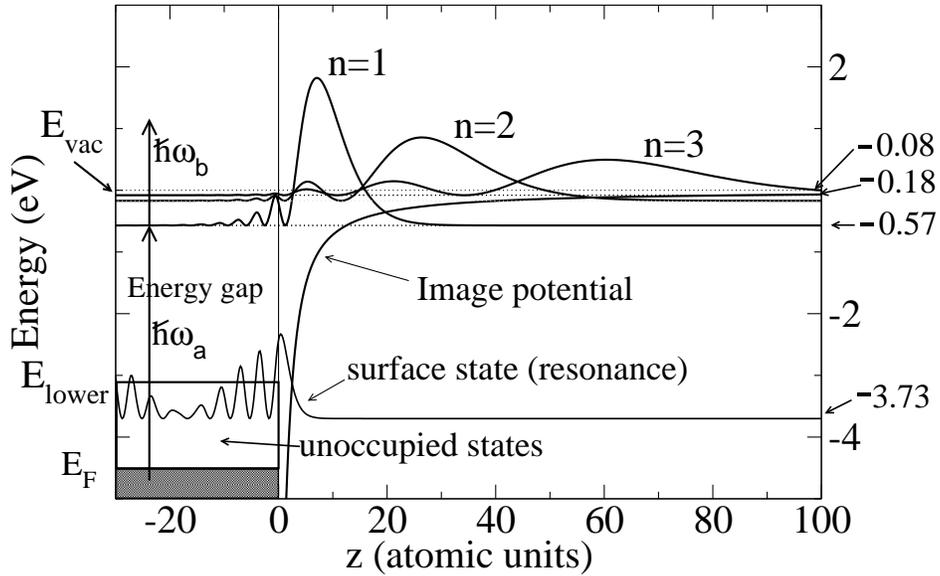}
\caption{\protect\small 
Schematic picture of a 2PPE experiment for a typical metallic surface
with a surface band gap close to the Fermi level. The energy 
$\hbar \omega_a$ corresponds to the pump photon and $\hbar \omega_b$
to the probe photon in 2PPE. On the right side we show the tail of the image 
potential, the densities of the surface state and first three image potential 
states of Cu(100) calculated with the model potential of Ref.~\cite{Chulkov2}, and their
corresponding binding energies.
By $E_\mathrm{F}$, $E_\mathrm{lower}$ and $E_\mathrm{vac}$ we denote the Fermi energy,   
the lower edge of the energy gap and the vacuum energy.}
\label{fig1}
\end{center}
\end{figure}

The presence of a metal surface creates electron states that do not exist in the
bulk metal. For example, an electron situated at a distance $z$ from the metal 
surface experiments an attractive force, $F(z)=-e^2/(2z)^2$, equivalent to that 
produced by its image charge situated at distance $z$ inside the metal.
For large $z$, the potential generated by this surface induced charge
approaches the classical image potential, $V(z)=-e^2/4z$\footnote{%
This expression corresponds to assuming an infinite repulsive barrier
at the surface. In the more general case we have $V(z)=-\frac{e^2}{4(z-z_\mathrm{im})}
\frac{\epsilon - 1}{\epsilon +1}$, where $z_\mathrm{im}$ is the position of the
image plane (that depends on the electronic and structural properties of the
surface), and $\epsilon$ is the static dielectric constant of the surface.
}.
When the metal has a surface band gap near the vacuum level, 
the electrons below the vacuum level are trapped between the well of the image 
potential and the surface barrier. These quantized states correspond to image 
potential states, and are spatially localized in the region in front of the surface
(see Fig.~\ref{fig1}). 
Metal image states typically have a large spatial extension, but barely penetrate the 
bulk metal. They form a Rydberg-like series with energies, $E_n$,
approximately given by\cite{Pedro}:
\begin{equation}
  E_n=-\frac{0.85}{(n+a)^2}\,[\mathrm{eV}]\,,
\end{equation}
where $n$ is a positive integer, and $0\leq a \leq 0.5$ represents the quantum defect 
(which depends on the position and width of the gap). 
The energy of an electron trapped in an image state and moving parallel to the surface 
with momentum $\bf {k_\parallel}$ is therefore:
\begin{equation}
  E=\frac{\hbar^2k_\parallel^2}{2m_e}+E_n
\end{equation}
In contrast to bulk states, the small overlap between surface states and bulk
states reduces considerably the inelastic electron scattering, leading to 
a long lifetime, $\tau_n$, of the image states, that scales asymptotically
with the quantum number $n$ as $\tau_n\propto n^3$\cite{Chulkov2}.
Due to this fact, these states are very interesting for the study of 
electron correlations. Furthermore, image states play an important role 
in the laser induced chemical control of reactivity
at metal surfaces. It is therefore not surprising the considerable number
of studies focusing on this subject (see Ref.~\cite{Chemphys} and references within).

Image potential states of different surfaces have been observed 
experimentally\cite{experiment}. A powerful technique to measure their 
energy and lifetime is two-photon photoemission spectroscopy 
(2PPE)\cite{Giesen}. Recently, and with the advent of ultrafast 
laser technology it became possible to perform time-resolved 
2PPE spectroscopy (TR2PPE), which allowed, with the technique of 
quantum beat spectroscopy\cite{Hofer,Klamroth}, the direct measurement
of the lifetime of image states, even for states with large 
quantum number $n$.
The 2PPE technique is depicted schematically in Fig.~\ref{fig1}.
One photon (pump), of energy $\hbar \omega_a$, excites an electron from an 
occupied state below the Fermi energy ($E_\mathrm{F}$) to an image-potential 
state of quantum number $n$; Then a second photon (probe), with energy 
$\hbar \omega_b$, ejects the electron out of the surface, above the vacuum 
energy. This electron has a kinetic energy $E_k=\hbar \omega_b - E_n$, that
is measured in a detector far away from the surface.
By varying the delay between the pump and probe pulses, the intensity of the 
2PPE signal as a function of the delay will reflect the evolution of the population of 
the state $n$. From this information it is then possible to extract the lifetime
of the state $n$.

Lifetimes of image states for different metal surfaces have been 
calculated in the framework of the self-energy formalism in the GW
approximation\cite{Osma,Chemphys}.
The results obtained happen to be in rather good agreement with experiments. 
For example, the calculated lifetime of the first three image potential states of 
the Cu(100) surface are 
$\tau_1^\mathrm{theor}=30$\,fs, 
$\tau_2^\mathrm{theor}=132$\,fs, and
$\tau_3^\mathrm{theor}=367$\,fs\cite{Chemphys}, while the experimental
values are
$\tau_1^\mathrm{exp}=40\pm6$\,fs, 
$\tau_2^{exp}=110\pm10$\,fs, and
$\tau_3^{exp}=300\pm15$\,fs\cite{Hofer},
while for the first image state of Cu(111) surface the theoretical lifetime is
$\tau_1^\mathrm{theor}=17.5$\,fs\cite{Chemphys} and the experimental is
$\tau_1^\mathrm{exp}=15\pm5$\,fs\cite{McNeil}.

In the present paper we present calculations on energy-resolved 2PPE 
spectra of Cu(100) and Cu(111) surfaces. Our theoretical approach is
based on the propagation of an electron wave-packet in an one-dimensional
model potential\cite{Chulkov2} simulating the Copper surface. The
finite electronic lifetimes of the image potential states are taken into 
account by an empirical self-energy term. The photoemission spectra are 
obtained through a geometrical scheme described in Ref.~\cite{2bepub}.
Note that in this approach the potential is fixed during the whole
simulations, i.e., the change of the electronic screening during the
excitation is not taken into account. This could be important for
short laser excitations as done in TR2PPE.

To assert the relevance of this
approximation we also performed time-dependent simulations of the Copper 
surface, but including the change of the Hartree potential due to the
electronic excitations.

\section{Method}
To model the Copper surface we used a one-dimensional slab of 48 Copper layers, 
surrounded by 290\,a.u. of vacuum on each side. The large portion of vacuum is 
necessary in order to describe the four first image potential wave functions
(see Fig.~\ref{fig1}). The bulk and image potentials are modeled by the 1D potential model
of Ref.~\cite{Chulkov2}. This model reproduces the position and the width of the 
energy gap, as well as the energy of the surface state and of the first image 
potential state. Furthermore, it provides a good description of the electronic 
structure of simple and noble metal surfaces\cite{Chulkov2}.
In this model the number of bulk states is discrete for each value of $q_\parallel$,
and is related to the number of layers in the simulation. We do not believe that
this is a major limitation, essentially because all the relevant physics occurs close to 
$E_\mathrm{F}$, while the remaining states mainly act as a polarizable background.
The electrons are allowed to move and interact only in the z-direction, 
perpendicular to the surface, and we assume a parabolic dispersion in the 
parallel direction. Within this approximation, the Hamiltonian describing such 
system can be written as
\begin{equation}
  \label{hamilt0}
  \hat{H}_0 = -\frac{d^2}{dz^2} + \hat{v}_\mathrm{model}\,.
\end{equation}
This operator is discretized in real-space using a grid-spacing of 0.2\,a.u,
which is sufficiently small to allow for a proper description of the 
relevant electronic states.

At $t=0$ we assume the system to be in the ground-state of the 
Hamiltonian~(\ref{hamilt0}). We then propagate this state with the 
new Hamiltonian
\begin{equation}
  \label{Hamiltonian}
  \hat{H}(t) = \hat{H}_0 + \hat{v}_\mathrm{laser}(z,t) + \hat{\Sigma}(z)\,,
\end{equation}
where $\hat{v}_\mathrm{laser}(z,t)$ describes a laser polarized in the
$z$ direction that reads, in the length gauge, 
$\hat{v}_\mathrm{laser}(z,t) = zE(t)$. The electric field is composed
of a pump and a probe pulses,
\begin{equation}
  E(t) = E_\mathrm{pump}(t)\cos(\omega_\mathrm{pump}t) +
    E_\mathrm{probe}(t)\cos(\omega_\mathrm{probe}t)\,.
\end{equation}
the functions $E_\mathrm{pump/probe}(t)$ are envelope functions of the form 
\begin{equation}
  E_i(t) = E_i\cos^2\left[\frac{\pi}{\sigma_i}(t-t_i)\right]
  \Theta(t-t_i+\sigma_i/2)\Theta(-t+t_i+\sigma_i/2)
  \,,
\end{equation}
where $\sigma_i$ is the width of the pulse and $t_i$ the center of the pulse.
For the pump pulse, $t_\mathrm{pump}$ is simply $\sigma_\mathrm{pump}/2$, 
while for the probe pulse $t_\mathrm{probe} = t_\mathrm{D} + \sigma_\mathrm{pump}/2$,
with $t_\mathrm{D}$ the delay time between the pump and probe lasers.
In order to take into account the finite lifetime of the image potential states
we add to the Hamiltonian~(\ref{hamilt0}) a time-independent self-energy operator 
of the form:
\begin{equation}
  \label{self}
  \Sigma(z) = -\mathrm{i}\sum_{k=1}^N\Gamma_k\vert n_k\rangle\langle n_k\vert
\end{equation}
where $\vert n_i\rangle$ are the image potential states and $\Gamma_k$ are 
the inverse experimental lifetimes of each state taken from Ref.\cite{Chemphys}.
Note that $\hat{H}(0)=\hat{H}_0$.
The time-propagation is performed in real-time following the method of 
Ref.\cite{octopus}. This propagation scheme preserves unitarity for a 
hermitian Hamiltonian, and has been proven very robust and stable in diverse
applications.

\section{Energy-resolved spectra}

\begin{figure}
\begin{center}
\includegraphics[width=0.7\textwidth,clip]{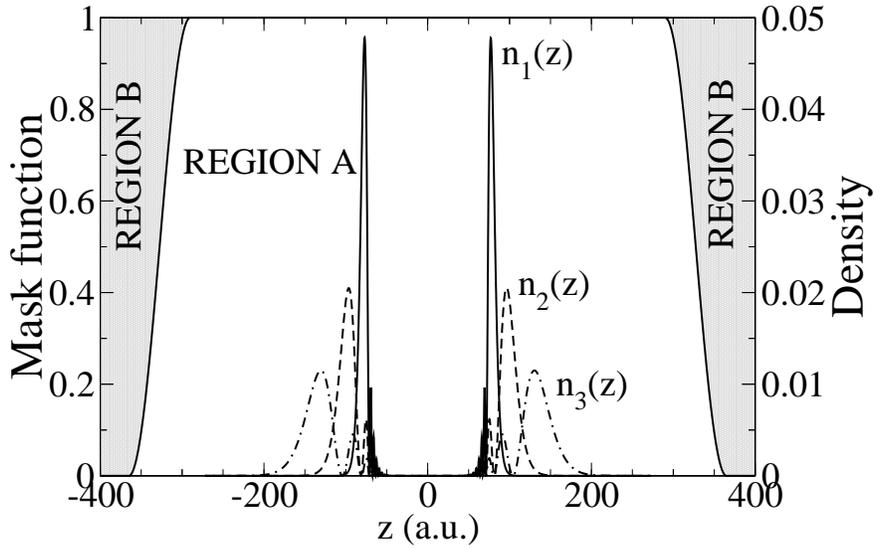}
\caption{\protect Typical shape of the mask function, $M$, used for dividing the 
space in two regions (see text for details). For illustrative purposes we also plot 
the densities of the first three image states of Cu(100).}
\label{fig2}
\end{center}
\end{figure}
In order to obtain the energy-resolved photo-electron spectra we follow the 
technique of Ref.\cite{2bepub} and divide the simulation box in two regions: 
one region, A, containing the slab and the region where the first
image potential states lie, and another, B, defined as the complement of A 
(see Fig.~\ref{fig2}). In region B the electrons are considered free outgoing 
particles, and are treated in momentum space. 
The separation between the two regions is achieved through a 
smooth masking function, $M$, defined as one in the interior of region A and zero 
outside. The method consists in evolving the wave-packet in the first region 
with the Hamiltonian~(\ref{Hamiltonian}), and mask the orbitals at each time-step.
The electrons that ``leave'' region A are then treated as free-particles and
accumulated in momentum space. In mathematical terms, the evolution is
performed in the following way: We start by propagating the wave-functions
in regions A and B using
\begin{eqnarray}
  \psi_k^\mathrm{A}(z,t+\Delta t) & = & M(z)\exp\left(-\mathrm{i}\hat H\Delta t\right)
    \psi_k^\mathrm{A}(z,t)
  \\ \nonumber
  \psi_k^\mathrm{B}(p,t+\Delta t) & = & 
    \exp\left\{-\mathrm{i}\frac{[p-A(t)]^2}{2}\Delta t\right\}\psi_k^\mathrm{B}(p,t) + 
    \tilde\psi_k^\mathrm{A}(p,t+\Delta t)
  \,,
\end{eqnarray}
where $\tilde \psi_k^A(p,t+\Delta t)$ is the Fourier transform of the 
part of the wave function $\psi_k^A$ that left region $A$ during the  
time step $\Delta t$, i.e,
\begin{equation}
  \tilde\psi_k^\mathrm{A}(p,t+\Delta t)=\int\!\!dz\:\exp(\mathrm{i}pz)
  [1-M(z)]\exp(-\mathrm{i}\hat H\Delta t)\psi_k^\mathrm{A}(z,t)
  \,,
\end{equation}
where $p$ denotes the momentum. If we wait long enough after that the laser has 
been turned off, region B will contain those electrons that were ionized.
The photo-electron spectra is then identified with:
\begin{equation}
  P(\sqrt{2mE}) = \sum_{k=1}^{N} \vert\psi_k^\mathrm{B}(p,t\rightarrow\infty)\vert^2
\end{equation}
where $N$ is the total number of electrons. This method can be derived from
the interpretation of the Wigner transform of the one-body density matrix
as a probability density\cite{2bepub}.

\begin{figure}
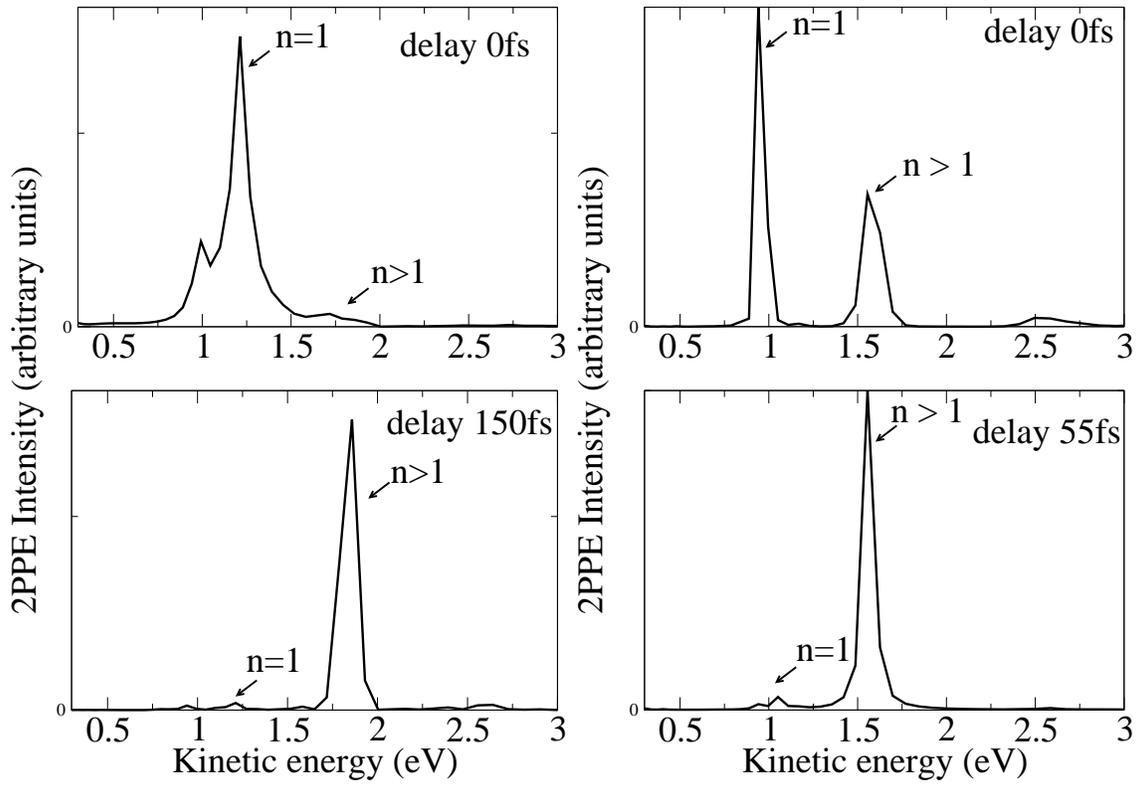

\begin{center}
\includegraphics[width=0.45\textwidth,angle=0,clip]{fig3.eps}
\includegraphics[width=0.45\textwidth,angle=0,clip]{fig4.eps}
\caption{\protect Photoelectron energy resolved spectra of a Cu(100) (left panel)
and a Cu(111) (right panel) surfaces for both zero and finite pump-probe delay.}
\label{fig3}
\end{center}
\end{figure}
In Fig.~\ref{fig3} we depict the energy resolved spectra of a Cu(100) 
and a Cu(111) surfaces for both zero and finite pump-probe delay. For the
(100) surface, the binding energies of the first image potential states 
are -0.57, -0.18, and -0.08\,eV (see Fig.~\ref{fig1}), while for the (111) surface the first
states are at -0.82, -0.22 and -0.009\,eV. Note that the second and
third image states of the Cu(111) surface are resonances.
Furthermore, the laser parameters used in these calculations were,
for the (100) surface, 
$\hbar \omega_\mathrm{pump} = 4.7$\,eV, $\hbar \omega_\mathrm{probe}=1.8$\,eV, 
$\sigma_\mathrm{pump}=95$\,fs, and $\sigma_\mathrm{probe}=54$\,fs, and for the (111) surface, 
$\hbar \omega_\mathrm{pump} = 5.1$\,eV, $\hbar \omega_\mathrm{probe}=1.8$\,eV, 
$\sigma_\mathrm{pump}=87$\,fs, and $\sigma_\mathrm{probe}=50$\,fs.
From these results we observe that the peaks corresponding to the first image 
potential state are obtained at the correct energy position. The signal due to 
the image states of higher quantum number cannot be distinguished due to the small 
difference in energy that cannot be resolved with our grid resolution.
We also note that, for zero delay, the broadening of the peak relative to the 
first image potential state is larger in the case of the Cu(100) surface than in 
the Cu(111). This is due to the fact that in the former case, the probe pulse 
can 'pump' low lying energy electrons from the bulk into energy levels near the gap
which can be then ejected from the surface by the pump pulse. As the Cu(111) surface
has a different electronic structure, such situation does not occur\cite{bands}.
Note that Cu(100) shows an intrinsic surface state resonance, while Cu(111) has
a surface state below the Fermi level. In the last case, the dominant transition
happens between the occupied surface state and the first image potential state.
This can be rationalized in terms of the larger spacial overlap between surface and
image states as compared to bulk states.

\section{Time-resolved spectra}
\begin{figure}
\begin{center}
\includegraphics[height=6.0cm,angle=0,clip]{fig5.eps}
\includegraphics[height=6.0cm,angle=0,clip]{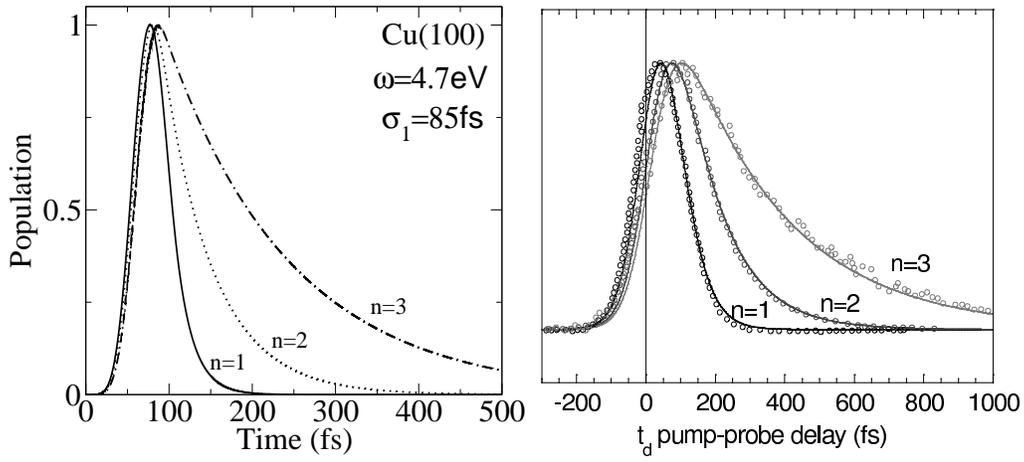}
\caption{\protect
Left: Population of the image states $n=1, 2$, and 3 as function of time.
The values of $\Gamma_{i=1,2,3}$ appearing in Eq.~(\ref{self}) 
were taken from Ref.~\cite{Chemphys},
and the maximum of each curve was normalized to one.
Right: 2PPE intensity for the $n=1, 2$, and 3 image states of Cu(100) as
function of the pump-probe delay. Reproduced from Ref.~\cite{Hofer}}
\label{fig5}
\end{center}
\end{figure}

The typical energy width of the laser pulses used in TR-2PPE varies 
between about 10 and 30\,meV. With this energy resolution it is possible
to excite separately each one of the first three image potential states of 
Cu(100). By following the TR-2PPE intensity in function of the pump-probe 
delay it is then possible to follow the evolution of the population of these
states. In Fig.\ref{fig5} we present calculations of the populations of the
first three image states of Cu(100). The curves were obtained by propagating
the system with the Hamiltonian~(\ref{Hamiltonian}). For comparison we also
show the experimental data from Ref.~\cite{Hofer}. Clearly, there is a good
qualitative agreement between the theoretical and experimental curves.

In these calculations, the electron-electron interaction was taken into account
by the model potential of Ref.~\cite{Chulkov2}. However, the variation 
of this term with time was mostly neglected. Part of the time-dependence
is taken indirectly into account through the self-energy operator, but
one can question how reasonable this approximation really is. In order
to answer to this concern, we performed simulations in which we allowed
the classical part of the electron-electron interaction (i.e. the Hartree
term) to change with time. This was achieved by adding to the 
Hamiltonian~(\ref{Hamiltonian}) a new term, $\delta v_\mathrm{Hartree}(z,t)$,
that corresponds to the variation of the Hartree potential due to
the change of electronic density in the direction perpendicular to the surface,
\begin{equation}
  \label{Hartree}
  \frac{\partial^2 \delta v_\mathrm{Hartree}(z,t)}{\partial z^2}=-4\pi\delta \rho(z,t)
  \,.
\end{equation}
where $\delta \rho(z,t) = \rho(z,t)-\rho(z,0)$.
Exchange and correlation effects were treated as before, i.e. through the 
self-energy term~(\ref{self}) used to simulate the lifetime of the states.
\begin{figure}
\begin{center}
\includegraphics[width=0.55\textwidth,clip]{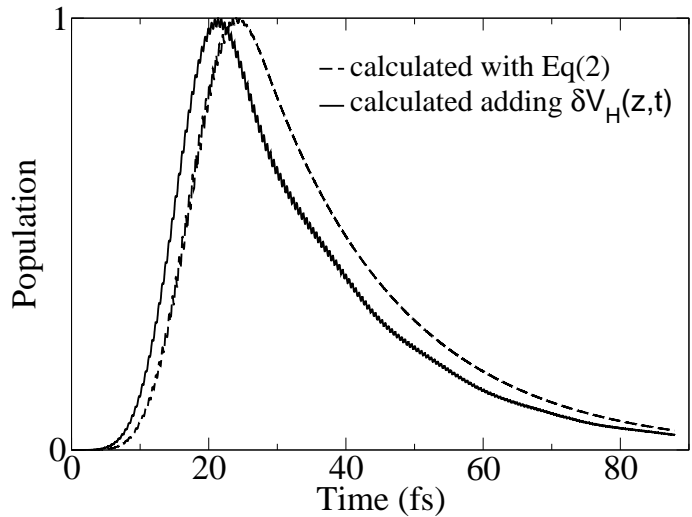}
\caption{\protect
Population of the first image potential state of Cu(100) calculated 
with and without the term of Eq.(\ref{Hartree}) for a 29\,fs laser pulse of 
energy $\hbar \omega= 4.2$\,eV. In both cases the maximum value of 
the population has been normalized to one. 
The difference in the lifetime for the two cases is $2\%$}
\label{fig7}
\end{center}
\end{figure}
The evolution of the population of the first image state with and without 
this term is shown in Fig~\ref{fig7}. We observe only a minor shift of the
peak to lower times from 24fs to 21.3fs and the lifetime is reduced  by around 2\%
when the term of Eq.(\ref{Hartree}) is included. 
This result validates previous studies that have neglected this effect\cite{Osma}. 

\section{Conclusions}

We have performed dynamical simulations in order to calculate
energy-resolved photoemission spectra of the Cu(100) and Cu(111) metal surfaces.
Using the same technique we also obtained time-resolved photoemission
spectra for individual excitations of the first three image potential states 
of Cu(100). This technique is quite general, and can be used to gain 
insight on the dynamics of image states with large quantum
number and when more than one eigenstate is excited coherently (the situation
in quantum beat spectroscopy).
Furthermore, we studied the influence of the variation of the Hartree potential 
during the excitation. This effects turns out to be small (around 2\%).

\section*{Acknowledgments} 
This work was supported by the Basque Country University, DGES and 
European Community under the research training network 
NANOPHASE (HPRN-CT-2000-00167). We wish to thank H. Appel, E.~K.~U. Gross
and P.~M. Echenique, for enlightening discussions and 
to the high-performance computing facilities of CEPBA where most of the 
calculations were done.


\begin{thebibliography}{99}
\bibitem{Pedro}
P.~M. Echenique and J.~B. Pendry, J. Phys. C  {\bf 11}, 2065 (1978);
P.~M. Echenique and J.~B. Pendry, Prog. Surf. Sci.  {\bf 32}, 111 (1990);
P.~M. Echenique, J.~M. Pitarke, E.~V. Chulkov, V.~M. Silkin  
J. El. Spec. Rel. Phen. {\bf 126}, 163 (2002).
\bibitem{Chulkov2}
E.~V. Chulkov, V.~M. Silkin and P.~M. Echenique,
Surf. Sci. {\bf 437}, 330 (1999).
\bibitem{Chemphys}
P.~M. Echenique, J.~M. Pitarke, E.~V. Chulkov and A. Rubio,
Chemical Physics {\bf 251}, 1 (2000);

\bibitem{experiment}
Chemical Physics {\bf 251}, 1 (2000); T. Fauster, W. Steinmann,  in: P. Halevi (Ed.),
Photonic Probes of Surfaces, Electromagnetic Waves: Recent Developments in Research Vol.2, 
Elsevier, Amsterdam, 1995.

\bibitem{Giesen}
K. Giesen et al, Phys. Rev. Lett. {\bf 55}, 300 (1985).

\bibitem{Hofer}
H. H\"ofer, I.~L. Shumay, C. Reu\ss, U. Thomann, W. Wallauer and T. Fauster,
Science {\bf 277}, 1480 (1997).

\bibitem{Klamroth}
T. Klamroth, P. Saalfrank and U. H\"ofer, Phys. Rev. B {\bf 64}, 035420 (2001).

\bibitem{Osma}
J. Osma, I. Sarria, E.~V. Chulkov, J.~M. Pitarke and P.~M. Echenique
Phys. Rev. B {\bf 59}, 10591 (1999); 
E.~V. Chulkov, J. Osma, I. Sarria, V.~M. Silkin and J.~M. Pitarke
Surf. Sci. {\bf 433}, 882 (1999);
I. Sarria, J. Osma, E.~V. Chulkov, J.~M. Pitarke and  P.~M. Echenique
Phys. Rev. B {\bf 60}, 11795 (1999);
E.~V. Chulkov, I. Sarria, V.~M. Silkin, J.~M. Pitarke and  P.~M. Echenique
Phys. Rev. Lett. {\bf 80}, 4947 (1998)
\bibitem{McNeil}
J.~D. McNeil, N.~H. Ge, C.~M. Wong, R.~E. Jordan, C.~B. Harris, 
Phys. Rev. Lett. {\bf 79}, 4645 (1997).

\bibitem{2bepub}
D. Varsano, M.~A.~L. Marques, H. Appel, E.~K.~U. Gross and A. Rubio,
{\it to be published}.


\bibitem{octopus} 
M.~A.~L. Marques, A. Castro, G.~F. Bertsch and A. Rubio, 
Comput. Phys. Commun. {\it in press} (2002).

\bibitem{bands}
For a schematical projection of the bulk band structure onto the (111) and (100)
surfaces of Cu see, for instance, Ref.\cite{Chemphys} pag. 21.  
 
\end{thebibliography}
\end{document}